# Regulated Polarization Degree of Upconversion Luminescence and Multiple Anti-Counterfeit Applications

**Dongping Wen, Ping Chen\*, Yi Liang, Xiaoming Mo, Caofeng Pan\***

D. Wen, Prof. Y. Liang, Prof. X. Mo, Prof. P. Chen\*, Prof. C. Pan\*
Center on Nanoenergy Research, Guangxi Key Laboratory for Relativistic Astrophysics, School of Physical Science and Technology, Guangxi University, Nanning 530004, China
Orcid.org/0000-0001-6875-9337; Email: chenping@gxu.edu.cn

Prof. C. Pan\*
CAS Center for Excellence in Nanoscience, Beijing Key Laboratory of Micronano Energy and Sensor, Beijing Institute of Nanoenergy and Nanosystems, Chinese Academy of Sciences, Beijing, 100140, China
Orcid.org/0000-0001-6327-9692; Email: cfpan@binn.cas.cn

Dongping Wen and Ping Chen have contributed equally to this work.

**Abstract** Polarization upconversion luminescence (PUCL) of lanthanide ($Ln^{3+}$) ions has been widely used in single particle tracking, microfluidics detection, three-dimensional displays, and so on. However, no effective strategy has been developed for modulating PUCL. Here, we report a strategy to regulate PUCL in $Ho^{3+}$-doped $NaYF_4$ single nanorods based on the number of upconversion photons. By constructing a multiphoton upconversion system for $Ho^{3+}$, we regulate the degree of polarization (DOP) of PUCL from 0.590 for two-photon luminescence to 0.929 for three-photon upconversion luminescence (UCL). Furthermore, our strategy is verified from cross-relaxation between $Ho^{3+}$ and $Yb^{3+}$, excitation wavelength, excitation power density, and local site symmetry. And this regulation strategy of PUCL has also been achieved in $Tm^{3+}$, where DOP is ranged from 0.233 for two-photon luminescence to 0.925 for four-photon UCL. Besides, multi-dimensional anti-counterfeiting display has been explored with PUCL. This work provides an effective strategy for regulating PUCL and also provides more opportunities for the development of polarization display, optical encoding, anti-counterfeiting, and integrated optical devices.

# Article Highlights

1. Novel excitation-polarization luminescence strategy regulated by number of upconversion photons.
2. The strategy enhanced $Ho^{3+}$ DOP from 0.166 to 0.929 and $Tm^{3+}$ DOP from 0.233 to 0.925, based on excited state population density.
3. Multi-dimensional anti-counterfeiting display and encryption are achieved (include words, pictures & labyrinth totem) via PUCL.

# 1 Introduction

Polarization upconversion luminescence (PUCL) of lanthanide ions ($Ln^{3+}$) is the anisotropic emission induced by the local site symmetry around $Ln^{3+}$ [1-3]. PUCL has developed applications in optical storage, biological imaging, polarization displays, and biotracking [4-9]. The adjustable polarized luminescence from $Ln^{3+}$ can provide a strong guarantee for biological probes and three-dimensional display [10-14]. In particular, highly PUCL has great potential in multiple anti-counterfeiting encryption and display [15-21]. However, due to the diversity of $Ln^{3+}$ ions and the complexity of the electronic structure of trivalent 4*f* ions [22], there is no effective control strategy for controlling the PUCL. The polarized luminescence of $Ln^{3+}$ is influenced not only by the local site symmetry of the crystal field (CF) but also many other factors, such as the direction of electric field of the excitation field, the energy transfer between $Ln^{3+}$, the concentration of activators, plasmonic nanomaterials with a specific structure, and so on [23-28]. In order to control the polarized luminescence of $Ln^{3+}$ from a single nanorod, attempts have been made to vary the concentration of sensitizers and activators [23,24]. However, it is still difficult to tune PUCL and achieve highly polarized luminescence from $Ln^{3+}$.

Site symmetry around $Ln^{3+}$ is essential for PUCL [1-3]. In the hexagonal phase of $NaYF_4$, point group symmetry is $C_s$ and energy states of $Ln^{3+}$ are represented by different irreducible $\Gamma_n$ [29]. Electrons are pumped from ground state (GS) to excited state (ES) $^{2s+1}L_J$, which is split into different Stark energy levels and presented irreducible ($\Gamma_n$). The ES undergoes a non-radiative (NR) or radiative transition and decays to the GS. This essential process for radiative transition is irreducible transition from $\Gamma_n$ (ES) → $\Gamma_m$ (GS) (m ≠ n) [30]. Some transition dipoles, such as σ and π configurations, exhibit excitation polarization dependence or emission anisotropy [3,31]. The polarization of the upconversion luminescence (UCL) originates from the mixture of different dipole transitions $\Gamma_n$ (ES) → $\Gamma_m$ (GS). Therefore, constructing an energy system for multiphoton upconversion can broaden the distribution of irreducible transitions $\Gamma_n$ → $\Gamma_m$ with various mixture states at multiple ESs. Here, we choose $Ho^{3+}$ and $Tm^{3+}$ as activators because they possess rich ladder-type energy levels and multi-color emissions, corresponding to multiple photon upconversion processes [32]. This brings more possibilities for our regulation of excitation polarization luminescence and extending the application to multiple anti-counterfeiting displays. Despite the promising performance of $Ho^{3+}$ and $Tm^{3+}$, the regulation of PUCL by the NR transitions therein remains a challenge.

In this work, we report a strategy to modulate the PUCL based on the multiphoton upconversion processes. The polarization of multiple emissions is regulated by the number of photons in the multiphoton upconversion process constructed in $Ho^{3+}$ ions, where the degree of polarization (DOP) of three-photon process being larger than that of two-photon process. With the exclusion of the CF, our mechanistic studies show that the multiphoton upconversion process can be pumped to a higher ES, which is less affected by NR transitions. The population density of higher ESs is lower, and the mixture of electron configuration for irreducible transitions $\Gamma_n$ (ES) → $\Gamma_m$ (GS) is fewer. That is, dipole orientations for irreducible transitions in higher ESs are not too much complicated. Thus, the proportion of dipole transitions in a similar direction is larger, resulting in a larger DOP of PUCL. The DOP exhibits an inverse relationship with the population density of ESs, which determines the strength of mixed irreducible transitions $\Gamma_n$ (ES) → $\Gamma_m$ (GS) and mixed dipole orientations. Therefore, we can tune the DOP of PUCL by controlling the number of photons in upconversion process. This regulatory strategy has been verified not only in $Ho^{3+}$ but also in $Tm^{3+}$. The DOP is regulated from 0.925 for four-photon process to 0.233 for two-photon process in NaYF$_4$:Tm,Yb single nanorods. This highly polarized and tunable UCL is applied in multi-dimensional anti-counterfeiting displays.

## 2 Results and discussion

2.1 Constructed multiphoton upconversion energy system for regulating PUCL

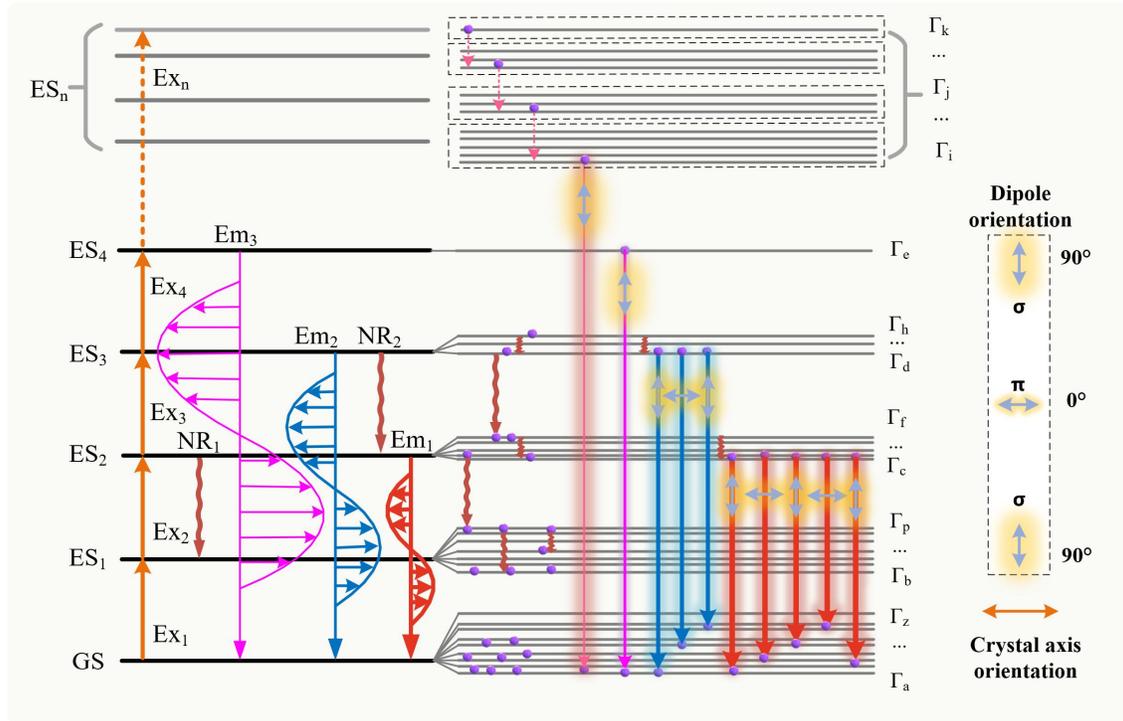

**Fig. 1** Schematic diagram of multiphoton enhancement of DOP of PUCL in Ln$^{3+}$ ions

In our work, we need to construct an energy system for multiphoton upconversion transitions to generate PUCL and realize polarization regulation. In the traditional multiphoton upconversion process by excited state absorption (ESA), ground state absorption (GSA) occurs once energy resonance of pump photon with the energy from the GS to the first ES$_1$ of Ln$^{3+}$. As a result, metastable state ES$_1$ is pumped, followed absorbing another pumped photon by ESA to generate $ES_2$. Then, $ES_2$ undergoes a radiative transition, producing two-photon UCL $Em_1$ (Fig. 1) [33]. Analogously, $ES_3$ gives rise to three-photon UCL, $Em_2$ and so on. According to:

$$N_2 \propto N_1^2 \text{ (i.e., } N_1 < 1), \tag{1}$$

where $N_1$ and $N_2$ are the population density of the $ES_1$ and $ES_2$, respectively [34]. More energy state $|i\rangle$ analogies. The energy state of $ES_4$ pumped by the four-photon upconversion process is greater than two-photon process, resulting in a lower population density ($N_4 < N_2$). Thus, under the same excitation power, the population density of ES with four-photon process is much smaller than that of two-photon process. Therefore, we can control the population density of $ES_i$ by the number of photons in upconversion processes.

In the hexagonal phase of $NaYF_4$, the $^{2s+1}L_J$ energy level of $Ln^{3+}$ can split into the 2J+1 energy level corresponding to irreducible $\Gamma_n$ symmetry [30, 31], because of the $C_s$ point group symmetry [29]. The irreducible transitions $\Gamma_d$ ($ES_3$) → $\Gamma_a$ (GS) (d ≠ a) generate UCL for the three-photon upconversion process (Fig. 1). Among these transitions, some transition dipoles exhibit excitation polarization dependence for dipole orientation. But the dipole orientations might vary from each other, such as σ and π polarization. And the lots of mixed dipole orientations reduce the probability of similar dipole orientations and results in low or canceled PUCL. Thus, the DOP of PUCL depends on mixed irreducible transitions $\Gamma_d$ ($ES_3$) → $\Gamma_a$ (GS) with different dipole orientations. Compared with the three-photon upconversion process, the population density of the $ES_4$ is lower because of four-photon upconversion process and the low energy transfer efficiency. And the mixing for different dipole orientations is reduced, which prefer to similar orientation and gives a higher DOP for PUCL. In contrast, the population density of the $ES_2$ is larger for two-photon upconversion process due to the less photons for UCL. As a result, the possibility of mixed different orientations of dipole transitions increases, and DOP of PUCL is smaller or even disappeared. Therefore, the number of upconversion photons influences the population density of ES and DOP of PUCL through the irreducible transition $\Gamma_i$ ($ES_n$) → $\Gamma_a$ (GS) with different dipole orientations (Fig. 1). However, the number of upconversion photons and the population density of $ES_i$ are also affected by NR transitions such as energy transfer, cross-relaxation, and excitation wavelength in the traditional upconversion process. We expect to control the DOP of PUCL through tuning the population density of $ES_i$ by the number of upconversion photons, accompanying with the interaction of NR transitions and the excitation wavelength.

## 2.2 Highly polarized NaYF$_4$:Ho upconversion luminescence nanorods

$Ho^{3+}$ is chosen first to realize the modulation of DOP by the number of upconversion photons for its excellent photochemical stability and ladder-type energy levels. NaYF$_4$:Ho nanorods (length ≈ 1460 ± 200 nm, diameter ≈ 108 ± 20 nm) are synthesized by a typical hydrothermal method (Fig. S1a,b), and all nanorods exhibited excellent uniformity in length and elemental distribution (Fig. S1c and Fig. 2b). Optical measurements of single nanorod are realized by home-built micro-region platform (Fig. 2a), where a half-wave plate is placed in the excitation path to change the direction of the electronic field of excitation light. Under the excitation of an 1150 nm laser, the UCL spectra (Fig. 2c) from NaYF$_4$:Ho single nanorod is obtained. Four emission peaks are presented, centered around 486 nm, 541 nm, 750 nm, and 648 nm, respectively, corresponding to the four transitions of $Ho^{3+}$ ions: $^5F_3 \rightarrow {}^5I_8$, $^5F_4/^5S_2 \rightarrow {}^5I_8$, $^5F_4/^5S_2 \rightarrow {}^5I_7$, and $^5F_5 \rightarrow {}^5I_8$, respectively. Then, PUCL spectra of NaYF$_4$:Ho single nanorod can be obtained (Fig. 2c) by changing the polarization direction of the excitation laser (Fig. 2a). Anisotropic absorption is first excluded because the vertical relationship between the orientation of polarized luminescence and anisotropic absorption (Fig. S2a,b and Fig. S1d). The luminescence intensities of the four transitions show the same trend with varying polarization angles. All of the transitions appear a period of 180° by fitting the luminescence intensities using:

$$y = y_0 + A * sin^2(\theta) \tag{2}$$

function (Fig. S1d). The DOP is defined as:

$$DOP = \frac{(I_{max} - I_{min})}{(I_{max} + I_{min})}, \tag{3}$$

where $I_{max}$ and $I_{min}$ are the maximum and minimum intensities of integrated UCL, respectively. The DOP for the $^5F_3 \rightarrow {}^5I_8$, $^5F_4/^5S_2 \rightarrow {}^5I_8$, $^5F_4/^5S_2 \rightarrow {}^5I_7$ and $^5F_5 \rightarrow {}^5I_8$ transitions are calculated to be 0.833, 0.828, 0.824, and 0.651, respectively. It is interesting that the DOPs of $^5F_3 \rightarrow {}^5I_8$, $^5F_4/^5S_2 \rightarrow {}^5I_8$, and $^5F_4/^5S_2 \rightarrow {}^5I_7$ transitions are close to each other, which are larger than the transition of $^5F_5 \rightarrow {}^5I_8$. According to the relationship between the DOP and the number of upconversion photons, this phenomenon implies that the number of upconversion photons for $^5F_3 \rightarrow {}^5I_8$, $^5F_4/^5S_2 \rightarrow {}^5I_8$, and $^5F_4/^5S_2 \rightarrow {}^5I_7$ transitions is consistent and larger than the number of photons for the $^5F_5 \rightarrow {}^5I_8$ transition.

Excitation power dependence of a single nanorod is carried out to assess the upcoversion process (Fig. 2d). The number of photons (*n*) required to populate the emitting state can be calculated by the relation of:

$$I_{em} \propto P^n, \tag{4}$$

where $I_{em}$ is the integral intensity of UCL and $P$ is the pump power of the laser [34]. The value of $n_1$ is 1.90, which corresponds to the $^5F_5 \rightarrow {}^5I_8$ transition, indicating that two photons are required to pump the $^5F_5$ state. The value of $n_2$, $n_3$, and $n_4$ for the three transitions of $^5F_4/^5S_2 \rightarrow$ $^5I_7$, $^5F_4/^5S_2 \rightarrow {}^5I_8$, and $^5F_3 \rightarrow {}^5I_8$ are 2.20, 2.32, and 2.44, respectively, suggesting the three-photon upconversion process. Thus, the DOPs of the transitions with three-photon pumping are larger than that of the transition with a two-photon process (Fig. 2e). This verifies our strategy to regulate the polarized luminescence based on the number of upconversion photons. The greater the number of upconversion photons, the smaller the population density of ESs and the higher the DOP of PUCL.

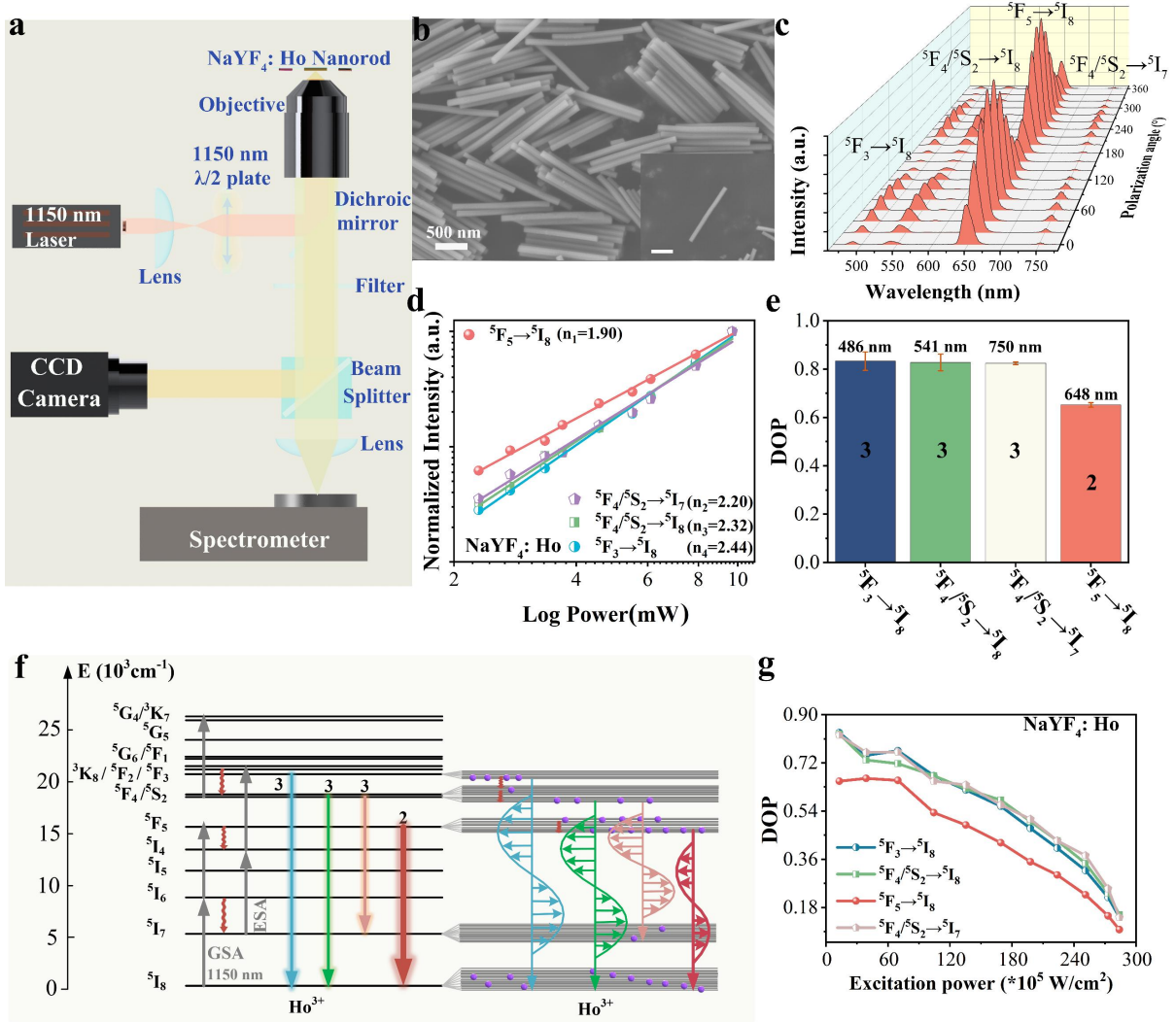

**Fig. 2** Regulated DOP of PUCL from $Ho^{3+}$ based on the number of upcoversion photon: **a** Schematic illuminated the UCL of $NaYF_4$:8% Ho single nanorods were collected by a home-built optical micro-region system; **b** Scanning electron microscope (SEM) image of $NaYF_4$:8% Ho nanorods and the inset is the single nanorod; **c** PUCL spectra of $NaYF_4$:8% Ho single nanorod excited at 1150 nm, the excitation power at 1150 nm for all tests was 11.8 mW at 100x objective, unless otherwise stated; **d** Upconversion emission intensity as a function of excitation power for $NaYF_4$:8% Ho single nanorod excited by 1150 nm; **e** Histogram illustrated the photon number connected DOPs of four transitions; **f** Schematic diagram of the multiphoton upconversion mechanism of $Ho^{3+}$ excited at 1150 nm, accompanying with the population density of $ES_i$ and DOPs of different transitions; **g** Dependence of the DOP on the pump power density of the excitation light

The DOPs of different transitions are regulated by the population density of $ES_i$. In the typical multiphoton upconversion transition process of $Ho^{3+}$ excited at 1150 nm (Fig. 2f), $^5F_3 \rightarrow {}^5I_8$, $^5F_4/^5S_2 \rightarrow {}^5I_8$, $^5F_4/^5S_2 \rightarrow {}^5I_7$ are three-photon upconversion processes and $^5F_5 \rightarrow {}^5I_8$ is a two-photon upconversion process, where the ESs of $^5F_4/^5S_2$ originate from the NR transition of the $^5F_3$ state. Any transition is a mixture of irreducible transitions $\Gamma_n$ (ES) $\rightarrow \Gamma_m$ (GS) with various dipole orientations [35]. Compared with the three-photon process, the two-photon process has a larger population density of ES. Thus, the mixing probability of different dipole orientations from two-photon process is much larger than those of the three-photon processes, resulting in a lower DOP. These phenomena are consistent with our proposed regulation strategy, indicating the DOP of PUCL can be adjusted by modulating the population density of ESs.

Furthermore, the population density of ES can be controlled by the pump power of the excited source, thus, the DOP of PUCL can be operated by the pump power density of the excitation light. Since the relationship:

$$N_i \propto P^n, \tag{5}$$

where $N_i$ is the population density of ESs and $P$ is excited power [36], the population density of ES $N_i$ will increase if enhancing the excited power $P$. Here, the statistics of DOP from a single nanorod on excited power are presented (Fig. 2g). It is found that the DOPs decrease for all of the upconversion transitions as the excitation power increases. But, the DOPs from three-photon transitions are still kept higher than that from two-photon transition at any power. Moreover, the DOPs are reduced to close to 0 for all the transitions, and PUCL are almost disappeared accompanying with much enhanced UCL intensity when the power is large enough. This is because the ESs of four transitions approach saturation at high power [36], inducing strong mixing of different dipole orientations from irreducible transitions $\Gamma_m$ (ES) → $\Gamma_n$ (GS) for any emission bands. Hence, the UCL are depolarized. The diminished trend of DOPs with increasing power indicates that the saturated population of ESs suppresses the PUCL from $Ln^{3+}$. Furthermore, the population density of ES can dominate the DOP, which verifies the polarized strategy regulated by the number of upconversion photons.

## 2.3 Cross-relaxation and the number of upconversion photons tune the DOPs of PUCL

The number of upconversion photons and the population density of ES are also affected by NR transitions such as cross-relaxation. The DOP strategy based on the number of upconversion photons is still effective with cross-relaxation. Here, we introduced cross-relaxation between $Yb^{3+}$ and $Ho^{3+}$ to monitor the DOPs from $Ho^{3+}$ (Fig. S3). The same emission peaks are detected under 1150 nm excitation (Fig. S4a), and PUCL also exhibits a periodic change of 180° with changing the polarization angle of the excitation light (Fig. 3a and Fig. S4b). On the one hand, the DOPs are 0.623 and 0.590 for the two-photon transitions $^5F_4/^5S_2 \rightarrow {}^5I_8$ and $^5F_4/^5S_2 \rightarrow {}^5I_7$, respectively (Fig. 3b,c). Both of the transitions come from the same ESs of $^5F_4/^5S_2$ and present similar DOPs. On the other hand, the DOPs are 0.800 and 0.929 for the three-photon transitions $^5F_5 \rightarrow {}^5I_8$ and $^5F_3 \rightarrow {}^5I_8$, respectively (Fig. 3b,c). Obviously, the DOPs of two-photon transitions are significantly smaller than those of three-photon transitions, which is consistent with the polarization strategy regulated by the number of upconversion photons. It is interesting that the $^5F_4/^5S_2 \rightarrow {}^5I_8$ and $^5F_4/^5S_2 \rightarrow {}^5I_7$ transitions are two-photon processes in NaYF$_4$:Yb,Ho single nanorod, while those are three-photon processes in NaYF$_4$:Ho single nanorod, indicating different pump paths for the same transitions. The anomalous upconversion pathways of the $^5F_4/^5S_2 \rightarrow {}^5I_8$, $^5F_4/^5S_2 \rightarrow {}^5I_7$, and $^5F_5 \rightarrow {}^5I_8$ transitions in nanorods are caused by cross-relaxation between $Yb^{3+}$ and $Ho^{3+}$ when excited by 1150 nm (Fig. 3d and Fig. S4c,d). These results indicate that the DOPs of PUCL are determined by the population density of ESs in the energy levels, rather than the energy level itself.

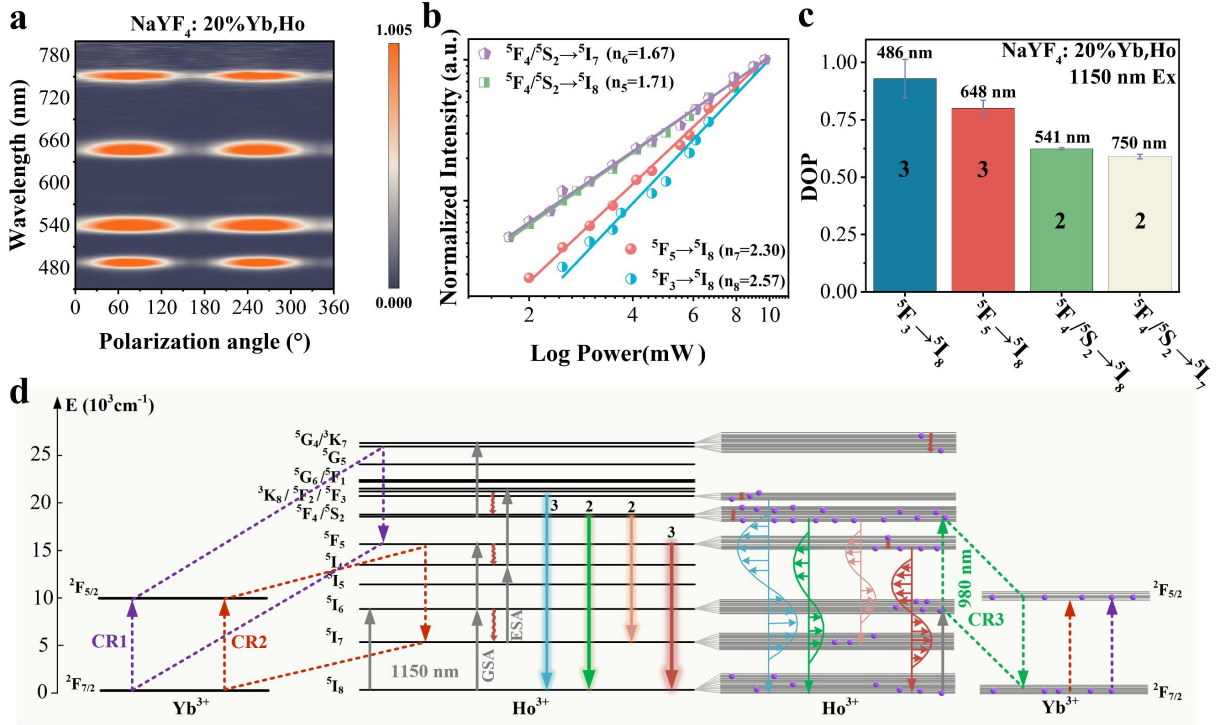

**Fig. 3** Tunable DOPs of PUCL from $Ho^{3+}$ with cross relaxation modulated the number of upconversion photons: **a** Matrix diagram of PUCL from NaYF$_4$:20% Yb,8% Ho single nanorod excited by 1150 nm; **b** Upconversion emission intensity as a function of excitation power for NaYF$_4$:20% Yb,8% Ho single nanorod excited by 1150 nm; **c** Histogram illustrated the photon number connecting DOPs of four transitions; **d** Schematic illustration of the multiphoton upconversion mechanism of NaYF$_4$:Yb,Ho excited at 1150 nm, accompanying with the population density of ES$_i$ and DOPs of different transitions

2.4 Excitation wavelength regulates DOPs of PUCL

The number of upconversion photons is correlated to the wavelength of excitation light. Generally speaking, for the same ES, the number of required upconversion photons with high energy is less than that with low energy because of the different pumping path. We changed the excitation wavelength to 975 nm to vary the pump path and regulate DOP (Fig. 4a and Fig. S5a,b). DOPs for two-photon transitions $^5F_4/^5S_2 \rightarrow {}^5I_8$ and $^5F_4/^5S_2 \rightarrow {}^5I_7$ are 0.166 and 0.155, respectively. And DOPs for three-photon process $^5F_5 \rightarrow {}^5I_8$ and $^5F_3 \rightarrow {}^5I_8$ transitions are 0.221 and 0.273, respectively (Fig. 4b,c). The greater the number of photons required, the higher the DOPs of UCL, which is consistent with our proposed strategy of controlling the DOP by the number of photons. In addition, DOP of $^5F_5 \rightarrow {}^5I_8$ transition is smaller than that of $^5F_3 \rightarrow {}^5I_8$ transition, although the same number of required photons for $^5F_5 \rightarrow {}^5I_8$ and $^5F_3 \rightarrow {}^5I_8$ transitions. The difference is originated from the various pump channels. The $^5F_5$ state comes from two paths. One way is the general two-photon upconversion process. The other way is the three-photon process with cross-relaxation (CR1): $^5G_4/^3K_7$ ($Ho^{3+}$) + $^2F_{7/2}$ ($Yb^{3+}$) $\rightarrow$ $^5F_5$ ($Ho^{3+}$) + $^2F_{5/2}$ ($Yb^{3+}$) (Fig. S5c). While $^5F_3$ state is pumped by three-photon absolutely. Thus, the population density of $^5F_5$ state is higher than $^5F_3$ state, resulting in a lower DOP for the $^5F_5 \rightarrow {}^5I_8$ transition.

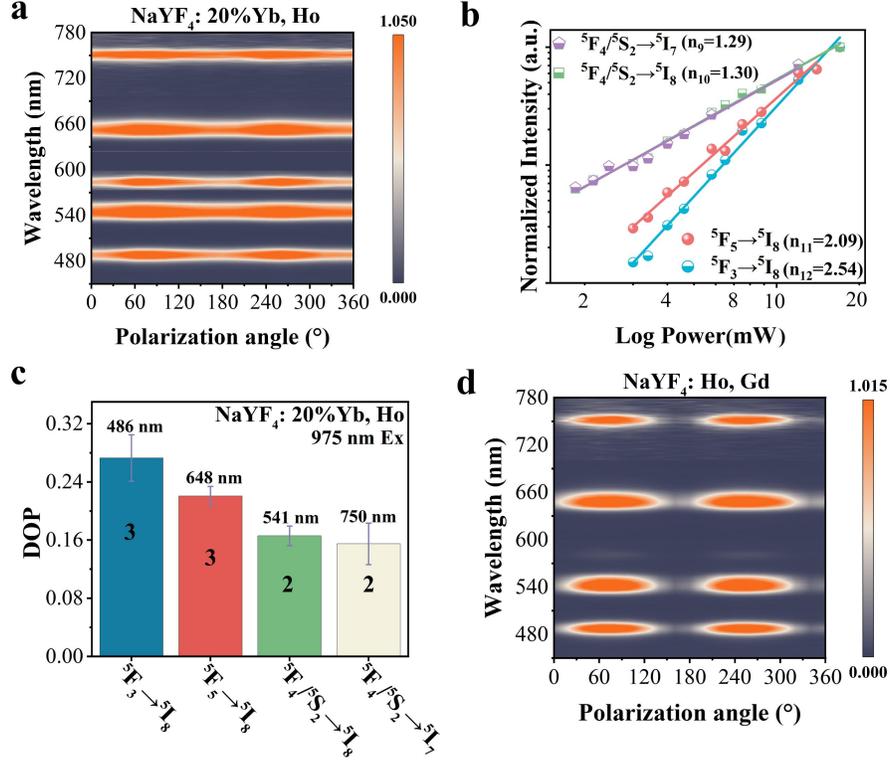

**Fig. 4** Excitation wavelength regulates the DOPs of PUCL synergistically: **a** Matrix diagram of UCL spectra from NaYF$_4$:20% Yb,8% Ho single nanorod excited at 975 nm, the excitation power at 975 nm for all tests was 680 μW at 100x objective, unless otherwise stated; **b** Upconversion emission intensity as a function of excitation power for NaYF$_4$:20% Yb,8% Ho single nanorod excited by 980 nm; **c** Histogram illustrated the photon number connected DOPs of four transitions; **d** Matrix diagram of UCL spectra from NaYF$_4$:10% Gd, Ho single nanorod excited at 1150 nm

The population density of ESs is also tuned by the synergistic effect of excitation wavelength and Yb$^{3+}$ doping based on the number of upconversion photons. When excited by 975 nm, the population density of ESs will have a significant increment because of the larger absorption cross of Yb$^{3+}$ and more efficient energy transfer from Yb$^{3+}$ to Ho$^{3+}$ [37,38]. The higher population density of ESs can induce a strong mixing of different dipole orientations and lead to a low DOP. Indeed, with the same number of upconversion photons, the DOPs are 0.221 and 0.273 for $^5F_5 \rightarrow {}^5I_8$ and $^5F_3 \rightarrow {}^5I_8$ transitions (Fig. 4c), respectively, in $Yb^{3+}$ co-doped single nanorod with 975 nm excited, while they are 0.800 and 0.929 in the same single nanorod with 1150 nm excited (Fig. 3c). The same situation exists for two-photon $^5F_4/^5S_2 \rightarrow {}^5I_8$ and $^5F_4/^5S_2 \rightarrow {}^5I_7$ transitions (Fig. 4c and Fig. 3c). The remarkable reduction of DOPs from $Yb^{3+}$ co-doped single nanorod reconfirms the regulated strategy of population density of ESs and the number of upconversion photons.

The regulated strategy based on the number of upconversion photons keeps working with the changing local site symmetry around activators. It has been reported that the bonding length and bonding angles between $Ln^{3+}$ and $F^-$ are altered with co-doped $Gd^{3+}$ into nanorods, which changes the local symmetry and polarized behavior of $Ln^{3+}$ [1,39]. In this section, we introduced a high concentration of $Gd^{3+}$ ions into $Ho^{3+}$-doped nanorods (Fig. 4d and Fig. S6a). The DOPs for $^5F_3 \rightarrow {}^5I_8$, $^5F_4/^5S_2 \rightarrow {}^5I_8$, and $^5F_4/^5S_2 \rightarrow {}^5I_7$ transitions are 0.867, 0.878, and 0.814 (Fig. S6c), which are three-photon upconversion processes (Fig. S6b). And the DOP for $^5F_5 \rightarrow {}^5I_8$ transition is 0.686 (Fig. S6c), which is a two-photon upconversion process (Fig. S6b). Notably, the DOPs of three-photon upconversion are higher than that of the two-photon process. Thus, the presented performance is in line with regulated strategy.

## 2.5 Tunable DOPs of PUCL from $Tm^{3+}$ based on the number of upconversion photons

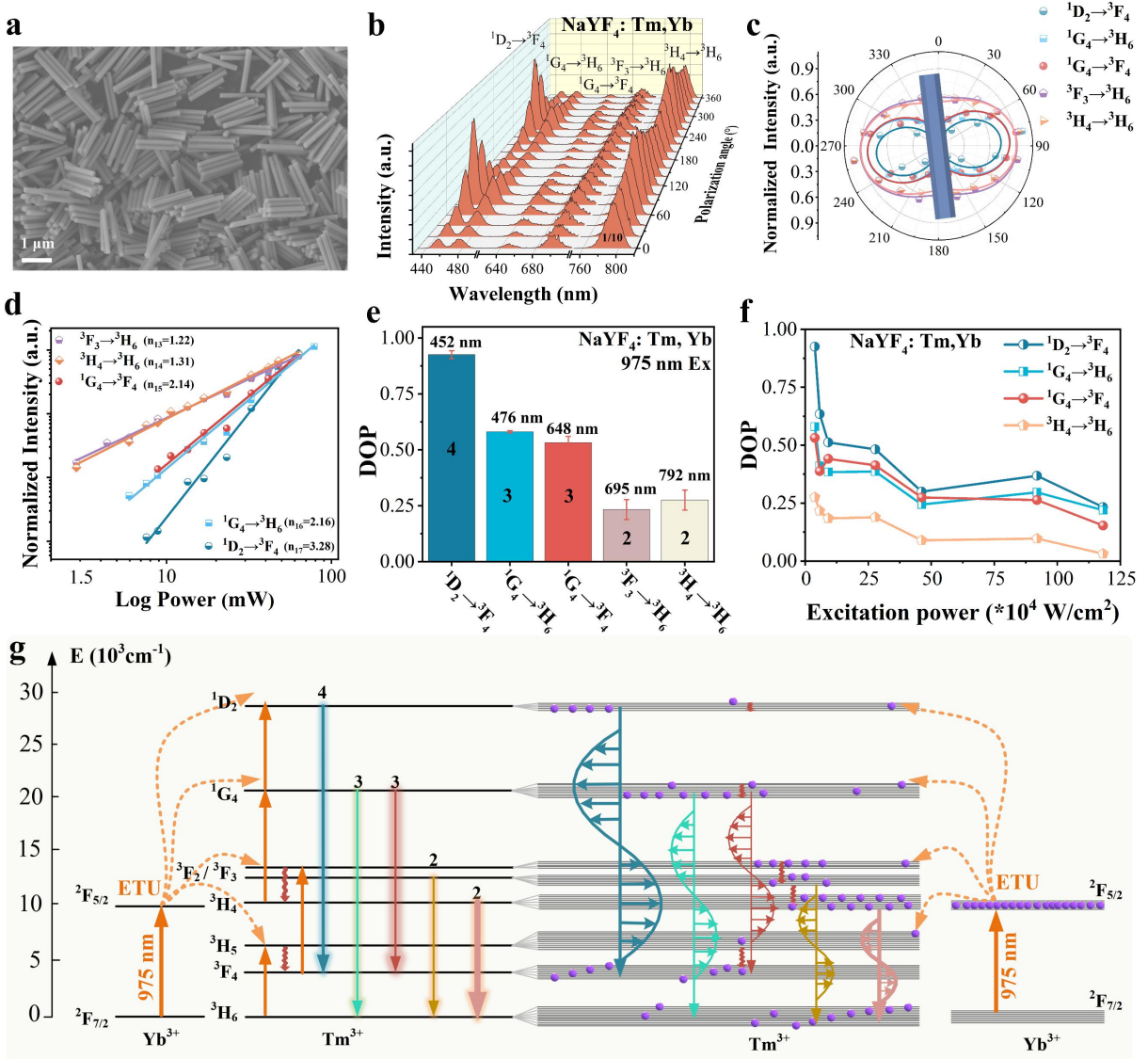

**Fig. 5** Controlled DOPs from NaYF$_4$:Tm,Yb single nanorods based on the number of upconversion photons: **a** SEM image of NaYF$_4$:5% Tm,20% Yb nanorods; **b** PUCL spectra from NaYF$_4$:Yb,Tm single nanorod excited at 975 nm, the excitation power at 975 nm for all tests was 680 μW at 100x objective, unless otherwise stated; **c** Polar plots the UCL integral intensities from $^1D_2 \to {}^3F_4$, $^1G_4 \to {}^3H_6$, $^1G_4 \to {}^3F_4$, $^3F_3 \to {}^3H_6$, $^3H_4 \to {}^3H_6$ transitions as a function of the excitation polarization angle; **d** Upconversion emission intensity as a function of excitation power for NaYF$_4$: 5% Tm,20% Yb single nanorod excited by 980

nm; **e** Histogram illustrated the photon number connected DOPs of five transitions; **f** DOPs of different transitions as a function of excitation power density; **g** Schematic diagram of the energy transfer from $Yb^{3+}$ to $Tm^{3+}$, accompanied with the DOPs based on the number of upconversion photons

The regulated strategy of DOPs based on the number of upconversion photons can be extended to $Tm^{3+}$ because of the excellent ladder-type energy levels to realize multiphoton upconversion [32]. The UCL spectra from $NaYF_4$:Tm,Yb single nanorod (Fig. 5a) present five emission bands upon excitation at 975 nm (Fig. 5b). The emission bands are centered around 452 nm, 476 nm, 648 nm, 695 nm, and 792 nm, corresponding to $^1D_2 \rightarrow {}^3F_4$, $^1G_4 \rightarrow {}^3H_6$, $^1G_4 \rightarrow {}^3F_4$, $^3F_3 \rightarrow {}^3H_6$, and $^3H_4 \rightarrow {}^3H_6$ transitions of $Tm^{3+}$, respectively. Then, the luminescence intensity was normalized by the deconvolution method and fitted by the function equation (2) (Fig. 5c). It is found that three kinds of DOPs are presented, corresponding to different upconversion modes. First is the two-photon upconversion path (Fig. 5d), the DOPs of $^3F_3 \rightarrow {}^3H_6$ and $^3H_4 \rightarrow {}^3H_6$ transitions are 0.233 and 0.275, respectively (Fig. 5e). The second is the three-photon upconversion way (Fig. 5d), the DOPs of $^1G_4 \rightarrow {}^3H_6$ and $^1G_4 \rightarrow {}^3F_4$ transitions are 0.580 and 0.531, respectively (Fig. 5e). The third is the four-photon upconversion process (Fig. 5d), the DOP of $^1D_2 \rightarrow {}^3F_4$ transition is 0.925 (Fig. 5e). Evidently, the DOP for the four-photon process is larger than that for three-photon process. Furthermore, the DOPs from three-photon process are much greater than those from two-photon process. Those phenomena are highly coincident with the population density of ESs controlled by upconversion photons. In other words, the more photons required in upconversion process, the less the population density of ES (Fig. 5g). As a result, the fewer mixing of different dipole orientations, the higher DOP of UCL. It is worth noting that a huge span of DOP from 0.925 to 0.233 is successfully achieved in $Tm^{3+}$, which is attributed to the large difference of population density in ESs based on four-photon and two-photon paths.

The modulation of DOPs from $Tm^{3+}$ UCL can also be realized by adjusting the population density in ESs by power density of the excitation light. The population density in ESs of $Tm^{3+}$ increases with growing pump power of the excitation light, leading to an obvious reduced trend of DOPs for all emissions (Fig. 5f). For the ES $^1D_2$, which originates from four- photon pump way (Fig. 5d,g), the population density rapidly increases with enhancing excitation power. Then, a mixture of irreducible transitions $\Gamma_n$ ($^1D_2$) → $\Gamma_m$ ($^3F_4$) with opposite dipole orientations is raised, resulting in a low proportion for similar dipole orientation and the small DOP of PUCL (Fig. 5g). This is analogous to the transitions of $^1G_4 \to {}^3H_6$, $^1G_4 \to {}^3F_4$, $^3F_3 \to {}^3H_6$, and $^3H_4 \to {}^3H_6$. Moreover, the reduced trend of DOP for the four-photon transition $^1D_2 \to {}^3F_4$ is more sensitive than for the three-photon transitions $^1G_4 \to {}^3H_6$ and $^1G_4 \to {}^3F_4$, and three-photon transitions are fast than two-photon transition $^3H_4 \to {}^3H_6$. Therefore, the upconversion photons number regulated population density of ESs can be feedbacked by the DOPs from single nanorods and verified in $Tm^{3+}$ and $Ho^{3+}$.

## 2.6 Multi-dimensional polarized anti-counterfeiting display and encryption

The PUCL of single nanorods is able to be applied in anti-counterfeiting displays with multi-dimensional encryption. Here, we introduce an encryption key by PUCL for anti-counterfeiting displays with full use of the low and high DOPs of $NaYF_4$:Yb,Ho nanorods upon excitation at 975 nm and 1150 nm, respectively. The low DOP excited by 975 nm can ensure the whole display of all the patterns, while the high DOP excited by 1150 nm is able to present the detail pictures, which provides an excellent opportunity for us to customize related patterns as polarized anti-counterfeiting displays. In detail, a rectangular groove is etched in silicon wafer as the pixel point for patterns by electron beam lithography (EBL) and reactive ion etching (RIE) (Fig. S7a). The width of the groove is less than the length of a single nanorod (Fig. S7b), thus, nanorods can only be filled into the grooves along rectangular groove (Fig. S7c,e). By software design, the rectangular groove can be arranged into any pattern with customizable size according to the designer's idea. Thus, we can consider the rectangular groove as a pixel point.

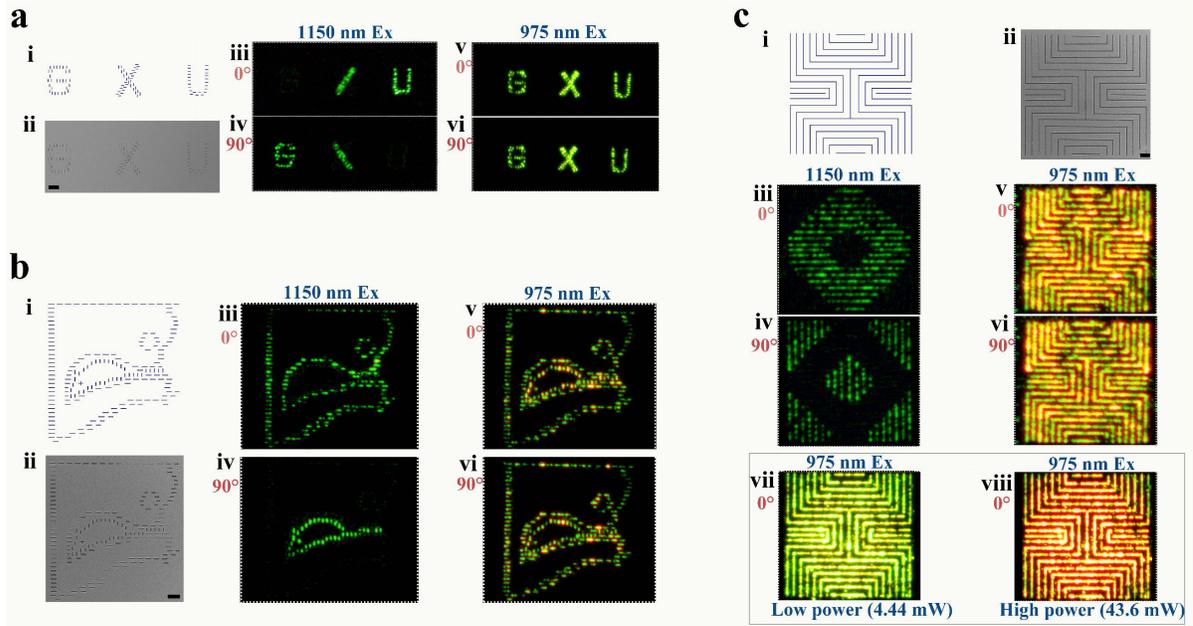

**Fig. 6** Excitation polarized display and optical anti-counterfeiting: **a** Design (i) and SEM (ii) images of "G", "X", and "U" letters composed of NaYF$_4$:Ho,Yb nanorods, where the pixels of "G" are arranged vertically; the left slashes of "X" ("/") are arranged vertically, and the right slashes ("\") are arranged horizontally; The pixels of "U" are arranged horizontally. Upconversion photoluminescence images of "G", "X", and "U" letters are excited by 1150 nm (iii-iv) or 975 nm (v-vi) with excitation polarized angle of 0° and 90°; **b** Design (i) and SEM (ii) images of "Cat" and "Mouse" patterns composed of NaYF$_4$:Ho,Yb nanorods, where "Cat" is composed of horizontal pixel arrangement and "Mouse" is composed of vertical pixel

arrangement; upconversion photoluminescence images of "Cat" and "Mouse" patterns excited by 1150 nm (iii-iv) or 975 nm (v-vi) with excitation polarized angles at 0° and 90°; **c** Design (i) and SEM (ii) images of "Labyrinth Totems" composed of NaYF$_4$:Ho,Yb nanorods, in which the vertical nanowire groove are arranged by vertical nanorods, and the horizontal nanowire groove are arranged by horizontal nanorods. Upconversion photoluminescence images of the "Labyrinth Totem" are excited by 1150 nm (iii-iv) or 975 nm (v-vi) with excitation polarized angle of 0° and 90°. Upconversion photoluminescence images of the "Labyrinth Totem" are excited by 975 nm with a low (vii) and high (viii) power, respectively. All scale bars are 10 μm. The excitation power at 1150 nm was 93 mW at the 10x objective, while the excitation power at 975 nm was 10.06 mW

"G", "X", and "U" letters were first etched and filled by NaYF$_4$:Yb,Ho nanorods (Fig. 6a). The "G" and "U" are arranged by vertical and horizontal nanorods, respectively. "X" is arrayed by vertical ("/") and horizontal ("\") nanorods, respectively (Fig. 6a-i,ii). Upon excitation at 1150 nm with a polarization angle of 0°, "U" and "/" of "X" letters are illuminated, while "\" of "X" and "G" letters are hidden (Fig. 6a-iii). When the angle was changed to 90°, "\" of "X" and "G" letters were displayed, while "U" and "/" of "X" letters disappeared (Fig. 6a-iv). At the same time, the whole letters "G", "X", and "U" can be completely shown under excitation at 975 nm due to the low DOP of nanorods (Fig. 6a-v and 6a-vi). Therefore, selective encryption and display can be realized by using different excitation wavelengths.

Graph encryption can be extended from the letter's encryption (Fig. 6b). The "Cat" and "Mouse" profiles are etched, where the "Cat" and "Mouse" are filled by horizontal and vertical nanorods, respectively (Fig. 6b-i,ii). We can see the "Cat" patterns when excitation is at 1150 nm and 0° of polarized angle (Fig. 6b-iii). While changing the polarized angle to 90°, the "Mouse" graph is presented (Fig. 6b-iv). In addition, when excited at 975 nm, the "Cat" and "Mouse" pictures are appeared simultaneously whenever the polarization angle of the excitation source (Fig. 6b-v and 6b-vi). These horizontal and vertical designs not only perform excellently in the encryption and hiding of words, but also present a brilliant performance in the anti-counterfeiting encryption of pictures.

Furthermore, high-polarized signal can be achieved in "Labyrinth Totem" composed of vertical and horizontal nanowire grooves (Fig. 6c-i,ii). The width of each nanowire groove is also smaller than the length of $NaYF_4$:Yb,Ho nanorods, thus, the nanorods can only be arrayed along the direction of nanowire grooves (Fig. S7d). Upon excitation at 1150 nm and 0° of polarized angle, a hollow diamond is presented (Fig. 6c-iii), which disappears when the excitation polarized angle is changed to 90° of (Fig. 6c-iv). Moreover, the whole "Labyrinth Totem" is illuminated in orange color when excited by 975 nm (Fig. 6c-v and 6c-vi). Besides, the green (Fig. 6c-vii) and red (Fig. 6c-viii) colors of the whole "Labyrinth Totem" emerged when excitation at 975 nm with a low and high power, respectively. Finally, five-fold anti-counterfeiting is achieved. Our design provides multiple guarantees for anti-counterfeiting displays and opens up infinite possibilities for polarized luminescence in designable patterns and multi-dimensional anti-counterfeiting displays.

## 3 Conclusion

In summary, we demonstrate a strategy to regulate the DOPs of PUCL from $Ln^{3+}$-doped single nanorods based on the number of upconversion photons. The core of this strategy is the population density of ESs tuned by the number of upconversion photons. Compared with the two-photon ESs, population density of ESs, which required much more photons, is lower. The lower population density generates a higher possibility of similar dipole orientation and a larger DOP of PUCL. The modulation of the DOPs of PUCL from $Ho^{3+}$ and $Tm^{3+}$ is realized by cross-relaxation, excitation wavelength, excitation power density, and changes in local site symmetry. The DOPs have been controlled from 0.233 of the two-photon upconversion process to 0.925 of the four-photon upconversion process in $Tm^{3+}$. Besides, taking advantage of the highly polarized nature of single nanorods, multi-dimensional anti-counterfeiting display and encryption in words, pictures, and labyrinth totem, have been realized. This regulated polarization strategy will promote the development of PUCL from $Ln^{3+}$-doped nanoparticles at the micro-nano scale. When the size of $Ln^{3+}$-doped nanoparticles decreases to several nanometers, non-radiative transitions should be considered due to the large specific surface area. The tunable polarization properties combined with multi-dimensional encryption methods enable potential applications in three-dimensional display, optical encoding, and many other emerging fields.

**Acknowledgements** The work is supported by National Natural Science Foundation of China (No. 11704081, 52125205, 52250398, U20A20166, 52192614 and 52003101 ), the Guangxi Natural Science Foundation (Nos. 2020GXNSFAA297182, 2020GXNSFAA297041, and 2017GXNSFBA19-8229), the special fund for "Guangxi Baigui Scholars", National key R&D program of China (2021YFB3200302 and 2021YFB3200304), Natural Science Foundation of Beijing Municipality (2222088), Shenzhen Science and Technology Innovation Program (Grant No. KQTD20170810105439418) and the Fundamental Research Funds for the Central Universities.

**Declarations**

**Conflict of interests** The authors declare that they have no conflict of interest.

**Supplementary Information** The online version contains supplementary material available at https://doi.org/10

Supporting Information

# Regulated Polarization Degree of Upconversion Luminescence and Multiple Anti-Counterfeit Applications

**Dongping Wen, Ping Chen\*, Yi Liang, Xiaoming Mo, Caofeng Pan\***

D. Wen, Prof. Y. Liang, Prof. X. Mo, Prof. P. Chen\*, Prof. C. Pan\*

Center on Nanoenergy Research, Guangxi Key Laboratory for Relativistic Astrophysics, School of Physical Science and Technology, Guangxi University, Nanning 530004, China

[Orcid.org/0000-0001-6875-9337](Orcid.org/0000-0001-6875-9337); Email: chenping@gxu.edu.cn

Prof. C. Pan\*

CAS Center for Excellence in Nanoscience, Beijing Key Laboratory of Micronano Energy and Sensor, Beijing Institute of Nanoenergy and Nanosystems, Chinese Academy of Sciences, Beijing, 100140, China

[Orcid.org/0000-0001-6327-9692](Orcid.org/0000-0001-6327-9692); Email: cfpan@binn.cas.cn

Dongping Wen and Ping Chen have contributed equally to this work.

**Contents**

**1 Experimental section**

**Fig. S1.** SEM images and Energy dispersive spectrum (EDS) mapping of NaYF$_4$:Ho nanorods

**Fig. S2.** Single-particle nanorod microscopic testing

**Fig. S3.** SEM images and EDS mapping of NaYF$_4$:Ho,Yb nanorods

**Fig. S4.** Spectroscopic testing of NaYF$_4$:8%Ho,20%Yb single nanorod excited by 1150 nm laser and cross-relaxation analysis between Ho-Yb

**Fig. S5.** PUCL spectra of NaYF$_4$:Ho,Yb nanorods excited by 975 nm laser

**Fig. S6.** PUCL spectra of NaYF$_4$:Ho,Gd nanorods excited by 1150 nm

**Fig. S7.** Template manufacturing process and SEM images of the rectangular groove and nanowire groove as pixel points

# 1 Experimental section

## 1.1 Materials and chemicals

All chemicals and solvents were obtained from commercial sources without further purification. NH$_4$F (99.99%), NaOH (99.99%), Y(NO$_3$)$_3$·6H$_2$O (99.99%), Yb(NO$_3$)$_3$·6H$_2$O (99.99%), Ho(NO$_3$)$_3$·6H$_2$O (99.99%), Tm(NO$_3$)$_3$·6H$_2$O (99.99%), and oleic acid (90%) were purchased from Sigma-Aldrich (St Louis, MO, USA). Ethanol, cyclohexane, and silicon wafers were purchased from Aladdin. E-Beam Resist PMMA 950K was purchased from ALLRESIST (German Tech Co., Ltd.).

## 1.2 Synthesis of NaYF$_4$ nanorods

Lanthanide-doped nanoparticles were synthesized as described in a previous report [1]. NaYF$_4$ nanorods were synthesized by a hydrothermal method. In the typical synthesis of NaYF$_4$ nanorods, NaOH (0.3 g) was first dissolved in deionized water (1.5 mL) by adding oleic acid (5 mL) and ethanol (5 mL) as additives under continuous stirring. Then, NH$_4$F (2 M, 1 mL) was added to form a turbid mixture. Subsequently, RE(NO$_3$)$_3$ (RE = Y, Yb, Ho, Gd or Tm) (0.2 M, 2 mL) was poured into the mixture. After being stirred for 20 minutes, the mixture was transferred to a 50 mL stainless steel Teflon-lined autoclave and kept at 220 °C for 12 hours in an oven. The nanorods were collected by centrifugation and washed with ethanol and cyclohexane three times. Finally, the nanorods were dried at 60 °C for 6 hours before use.

## 1.3 Field emission scanning electron microscope characterization

SEM images were obtained from a ZEISS Sigma-500 field emission scanning electron microscope. Energy dispersive spectrum (EDS) characterization was performed using a ZEISS Sigma-500 and an EDS detector (Oxford X-MaxN20) operated at an acceleration voltage of 15 kV.

**1.4 Testing of polarized upconversion luminescence from single nanorod**

To measure the PUCL from lanthanide-doped $NaYF_4$ single nanorods, a home-built optical microscope system (Fig. 2a) was used: an external optical path was used to introduce 1150 nm (or 975 nm) linearly polarized light as the excited light source into an Olympus IX73 inverted microscope, which was coupled with an Edinburgh FLS1000 spectrometer. A half-wave plate at 1150 nm (or 975 nm) was placed in the path of the excited light. Photoluminescence spectra were recorded at room temperature using a FLS1000 spectrometer (Edinburgh Instruments) equipped with an PMT-900 photon-counting photomultiplier tube, and the PUCL spectra were obtained by rotating the half-wave plate.

**1.5 Display template manufacturing and nanorods assembly**

The positive e-beam resist of polymethyl methacrylate (PMMA) was spun onto the substrate of a silicon wafer with 300 nm silicon oxide at 4000 rpm for 60 s. Then the substrate was baked at 180 °C for 2 minutes. Following etched by EBL (RAITH ELPHY & ZEISS-SEM) at 10 KV acceleration voltage and 80 pA beam current. The etched patterns were immersed in a developer solution of methyl isobutyl ketone and isopropanol (IPA) (1:3) for 30 seconds and then flushed with IPA to present the patterns. Then, the patterns were etched by RIE (TEGAL 903E). Finally, the patterns with the pixel point of the rectangle groove were built. The nanorods were assembled in rectangular grooves by polydimethylsiloxane (PDMS). The nanorods, dispersed in ethanol, were dripped on the substrate, which was dragged across the surface several times using PDMS at a speed of 0.5 mm/s (Fig. S7e).

**1.6 The definition of laser spot diameter**

Upconversion luminescence was collected using a home-built optical microscope system, as shown in Fig. 2a. A 1150 nm or 975 nm laser was focused on a sample located at the surface of a microscope slide using a microscope objective with a high numerical aperture (NA = 1.3). The laser spot diameter was calculated by $d = (1.22 * \lambda) / NA$, where $\lambda$ is the excitation wavelength and $d$ is the laser spot diameter [2,3]. Therefore, the spot diameter of the 1150 nm excitation beam was estimated to be approximately 1080 nm in diameter. The length of nanorods is larger than the excitation spot diameter (1080 nm).

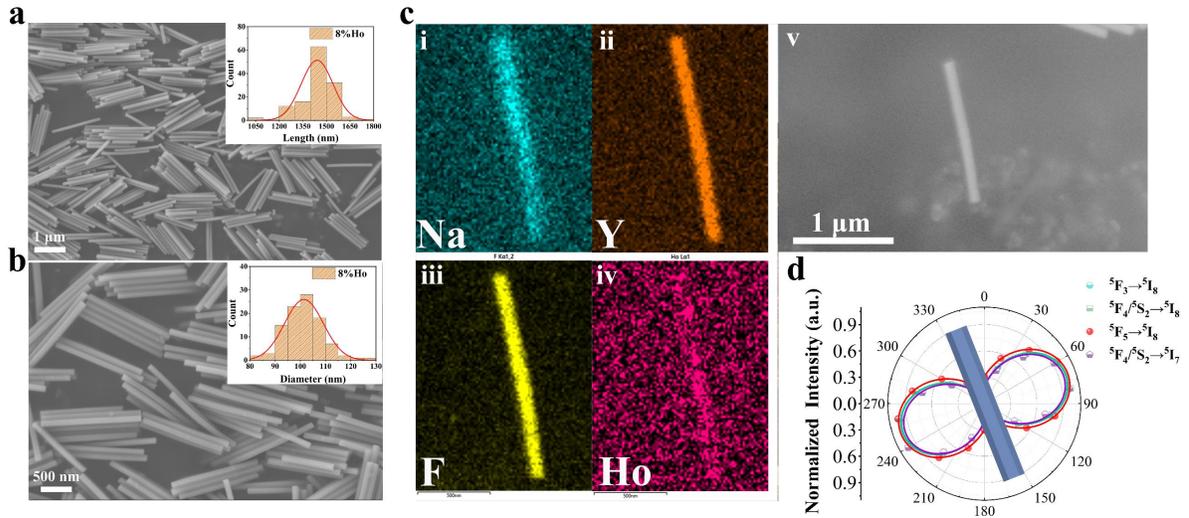

**Fig. S1** SEM images and Energy dispersive spectrum (EDS) mapping of NaYF$_4$:Ho nanorods: **a-b** SEM images of NaYF$_4$: 8% Ho nanorod, the insets are histogram showing the size in (**a**) *c* and (**b**) *a* axis; **c** EDS mapping of NaYF$_4$:Ho single nanorod; **d** Polar plots the upconversion luminescence integral intensities from $^5F_3 \rightarrow {}^5I_8$, $^5F_4/^5S_2 \rightarrow {}^5I_8$, $^5F_4/^5S_2 \rightarrow {}^5I_7$, $^5F_5 \rightarrow {}^5I_8$ transitions as a function of the excitation polarization angle, including the diagram of the relationship between the orientation of nanorods and the intensity of excited polarized luminescence

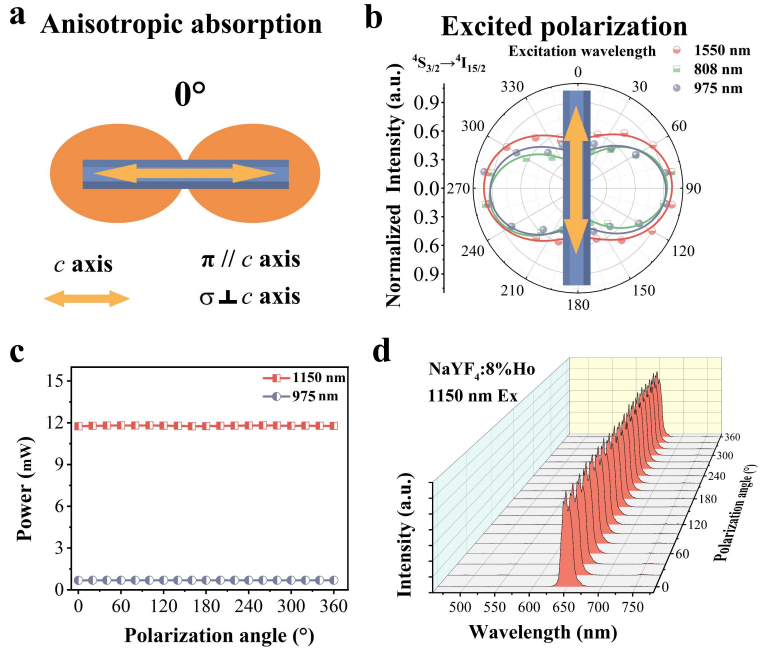

**Fig. S2** Single-Particle Nanorod Microscopic Testing: **a** Schematic diagram illustrates the parallel relationship of *c* axis of nanorod and polarized orientation originated from anisotropic absorption of β-NaYF$_4$:Er nanorods at 808 nm, 975 nm, and 1550 nm; **b** Schematic diagram illustrates vertical relationship between the *c* axis of nanorods and the polarized upconversion luminescence (PUCL) of the measured β-NaYF$_4$:Er nanorod under excitation at 808 nm, 975 nm, and 1550 nm, respectively; **c** Output power from microscope objective of excitation laser with changing excitation polarization angle at 1150 nm and 975 nm, respectively; **d** Upconversion luminescence spectra of the NaYF$_4$: 8%Ho nanoparticles with rotating the polarizer after the dichroic mirror in probe light path

The influence of anisotropic absorption on the excitation polarization response is excluded by measuring the orientation relationship between PUCL and anisotropic absorption. It has been reported that the intensity of anisotropic absorption from *c* axis is much stronger than that from the direction perpendicular to *c* axis at 1550 nm, 975 nm, and 808 nm in β-NaGdF$_4$: Er hexagonal crystal [4], whose structure is similar to β-NaYF$_4$: Er. Therefore, the excitation polarization luminescence should be along *c* axis if anisotropic absorption is the main reason (Fig. S2a). Actually, all of the polarized directions of upconversion luminescence from Er$^{3+}$ is perpendicular to *c* axis when excited at 1550 nm, 975 nm, and 808 nm (Fig. S2b), respectively, indicating non-leading role of anisotropic absorption.

    The reflection and transmission of dichroic mirror and other optical elements on excitation polarization response are eliminated. On one hand, the output power from the microscope objective was measured with rotating the half-wave plate from 0°-180° (Fig. S2c). The power output from microscope objective of both 1150 nm and 975 nm lasers remained almost unchanged with rotating the half-wave plate from 0°-180°. Thus, the reflection of dichroic mirror has little effect on excitation polarization response. On the other hand, to exclude the influence of transmission of dichroic mirror on excitation polarization response, the luminescence of NaYF$_4$: 8%Ho nanoparticles have been measured. The emission spectra did not change with rotating the polarizer after the dichroic mirror in probe light path. Because the beam diameter (1080 nm) is much larger than the nanoparticles (20-40 nm) and the random orientation of nanoparticles, no polarized luminescence should be observed, which is consistent with our experiment (Fig. S2d). Thus, the transmission of dichroic mirror has no obvious effect on excitation polarization response. Therefore, excitation-angle dependent reflection and transmission of dichroic mirror are not the important factors on our experiments and results.

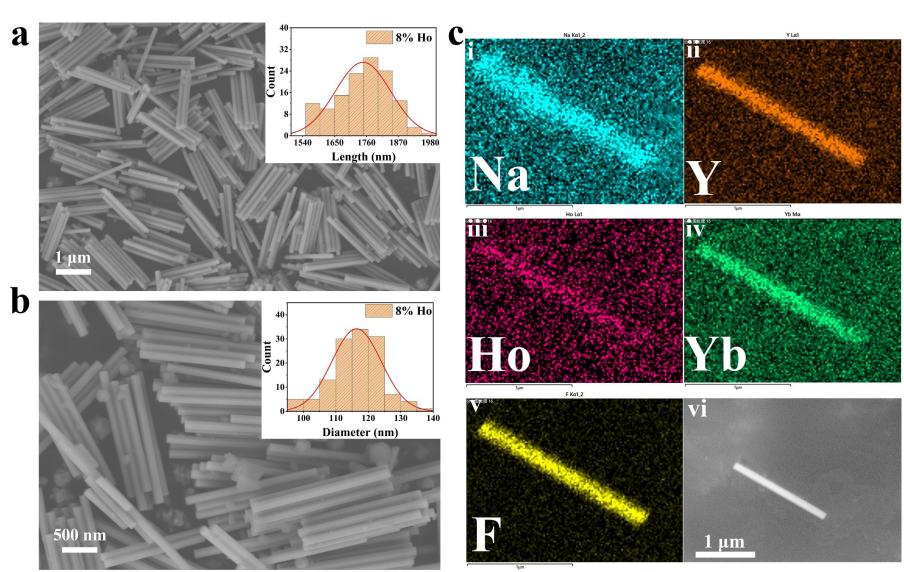

**Fig. S3** SEM images and EDS mapping of NaYF$_4$:Ho,Yb nanorods: **a-b** SEM images of NaYF$_4$: 8% Ho, 20% Yb nanorod. Insets are size histograms on the (**a**) *c* and (**b**) *a* axis; **c** EDS mapping of NaYF$_4$: 8% Ho, 20% Yb single nanorod

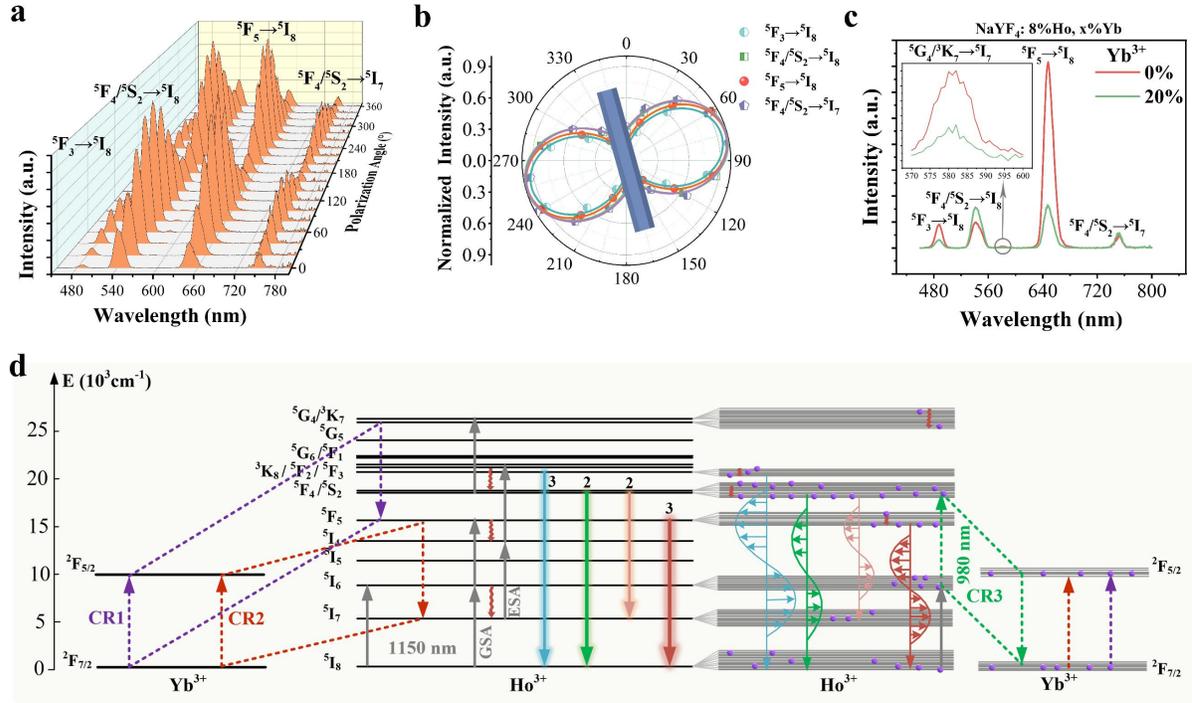

**Fig. S4** Spectroscopic testing of NaYF$_4$:8%Ho,20%Yb single nanorod excited by 1150 nm laser and cross-relaxation analysis between Ho-Yb: **a** PUCL spectra from NaYF$_4$: 20% Yb, 8% Ho single nanorod; **b** Polar plots upconversion luminescence integral intensities as a function of the excitation polarization angle from NaYF$_4$: 20% Yb, 8% Ho single nanorods; **c** UCL spectra of $^5F_4/^5S_2 \rightarrow {}^5I_8$, $^5F_4/^5S_2 \rightarrow {}^5I_7$, $^5F_5 \rightarrow {}^5I_8$ and $^5F_3 \rightarrow {}^5I_8$ at 0% Yb$^{3+}$ and 20%Yb$^{3+}$ concentrations excited at 1150 nm, the inset is UCL spectra of $^5G_4/^3K_7 \rightarrow {}^5I_7$; **d** Schematic illustration of the multiphoton upconversion mechanism of NaYF$_4$:Yb,Ho excited at 1150 nm, accompanying with the population density of ES$_i$ and DOPs of different transitions

The cross-relaxation (CR) of Ho$^{3+}$ and Yb$^{3+}$ causes the abnormal photon number in NaYF$_4$: Yb, Ho. When introducing Yb$^{3+}$ into Ho$^{3+}$-doped NaYF$_4$ nanorods, three kinds of cross-relaxation (CR1, CR2 and CR3) between Yb$^{3+}$ and Ho$^{3+}$ occur (Fig. S4d).

CR1 is $^5G_4/^3K_7$ (Ho$^{3+}$) + $^2F_{7/2}$ (Yb$^{3+}$) → $^5F_5$ (Ho$^{3+}$) + $^2F_{5/2}$ (Yb$^{3+}$). When introducing Yb$^{3+}$ into nanorods, luminescence intensities from the $^5G_4/^3K_7$ → $^5I_7$ transition are decreased and even disappear at 20%Yb$^{3+}$ concentration (Fig. S4c and S4d), suggesting CR1 between Yb$^{3+}$ and Ho$^{3+}$ [5].

CR2 is $^5F_5$ (Ho$^{3+}$) + $^2F_{7/2}$ (Yb$^{3+}$) → $^5I_7$ (Ho$^{3+}$) + $^2F_{5/2}$ (Yb$^{3+}$). Once introducing Yb$^{3+}$ into nanorods, the luminescence intensity from $^5F_5$ → $^5I_8$ is largely reduced (Fig. S4c and S4d), indicating the CR2 between Yb$^{3+}$ and Ho$^{3+}$ [5].

CR3 is $^5I_6$ (Ho$^{3+}$) + $^2F_{5/2}$ (Yb$^{3+}$) → $^5F_4/^5S_2$ (Ho$^{3+}$) + $^2F_{7/2}$ (Yb$^{3+}$). Accompanied by the introduction of Yb$^{3+}$ ions, luminescence intensities from $^5F_4/^5S_2$ → $^5I_8$ and $^5F_4/^5S_2$ → $^5I_7$ transitions are enhanced (Fig. S4c and S4d), suggesting CR3 between Yb$^{3+}$ and Ho$^{3+}$ [5].

The upconversion paths for $^5F_5$ → $^5I_8$ transition are originated from two way. First, the normal two-photon process is to populate $^5F_5$ state and followed by the $^5F_5$ → $^5I_8$ transition, which is reported in other literatures [5-9]. Second, the CR1 between Yb$^{3+}$ and Ho$^{3+}$ is the other path to populate $^5F_5$ state, following the $^5F_5$ → $^5I_8$ transition [5]. For the second way, $^5G_4/^3K_7$ states is originated from four-photon process [9]. Thus, the upconversion paths for $^5F_5$ → $^5I_8$ transition are the mixing processes of two-photon and four photon upconversion, presenting 2.3 for $n_7$ value in excitation power dependence of emission dynamics (Fig. 3b) and a three-photon-like upconversion process.

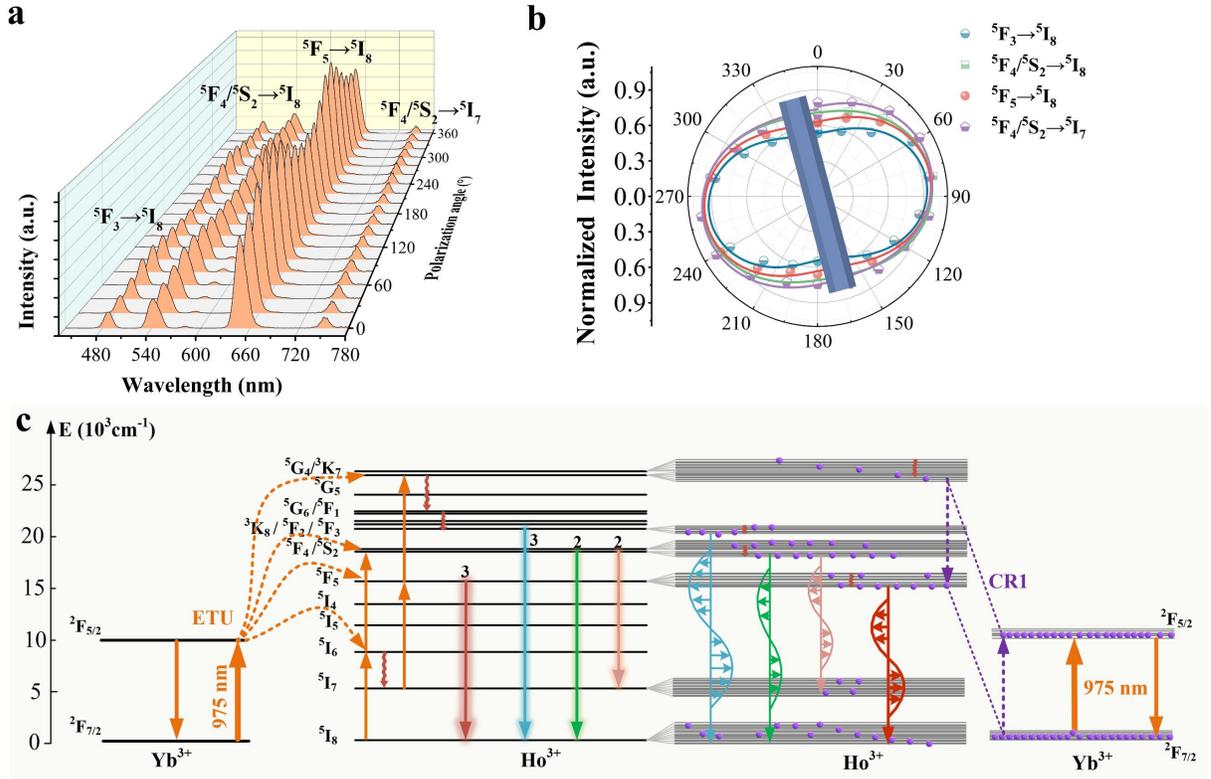

**Fig. S5** PUCL spectra of NaYF$_4$:Ho,Yb nanorods excited by 975 nm laser: **a** PUCL spectra from NaYF$_4$:8%Ho,20%Yb single nanorods; **b** Polar plots upconversion luminescence integral intensities as a function of the excitation polarization angle from NaYF$_4$:8%Ho,20%Yb single nanorods; **c** Schematic illustration of the multiphoton upconversion mechanism of NaYF$_4$:Yb,Ho excited at 975 nm, accompanying with DOPs regulated by the number of upconversion photons

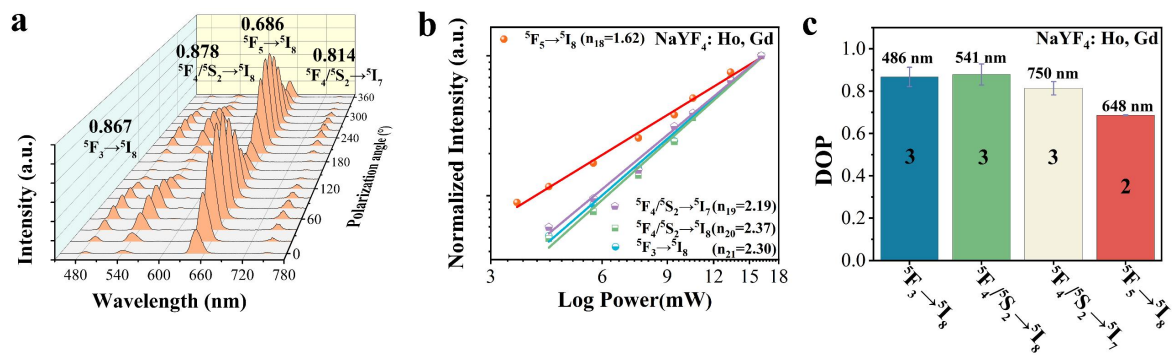

**Fig. S6** PUCL spectra of NaYF$_4$:Ho,10%Gd nanorods excited by 1150 nm: **a** PUCL spectra of NaYF$_4$:Ho,10%Gd single nanorod with excited polarization angles of 1150 nm from 0° to 360°; **b** Upconversion emission intensity as a function of excitation power for NaYF$_4$:10%Gd,Ho single nanorod excited by 1150 nm; **c** Histogram illustrated the photon number connected DOPs of four transitions from Gd$^{3+}$ co-doped single nanorod excited at 1150 nm

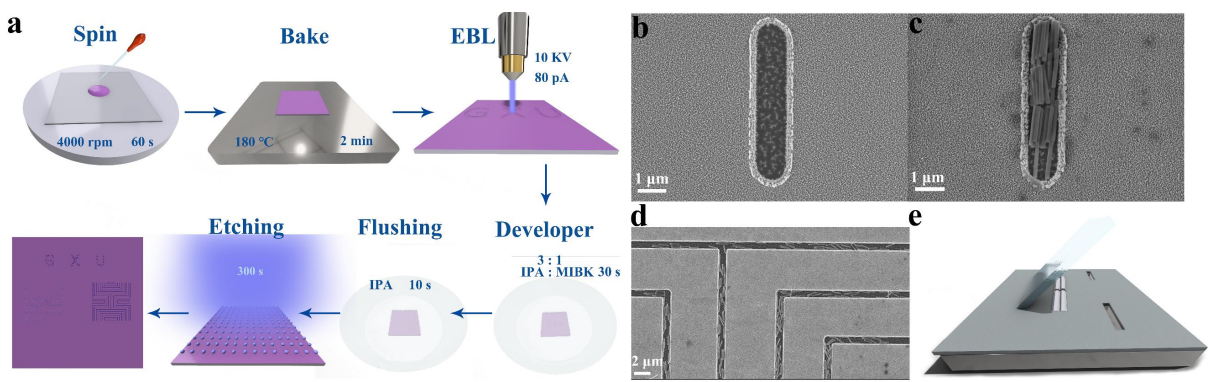

**Fig. S7** Template manufacturing process and SEM images of the rectangular groove and nanowire groove as pixel points: **a** Schematic diagram of the template manufacturing process; **b** SEM image of a rectangular groove with etching by EBL and RIE. **c** SEM image of rectangular groove filled by NaYF$_4$:Yb,Ho nanorods; **d** SEM image of nanowire groove are filled by NaYF$_4$:Yb,Ho nanorods. **e** Demonstration of the method of assembling nanorods by PDMS